\documentclass[twocolumn]{aastex62}
\usepackage{subfigure}

\received{December 17, 2018}
\revised{January 24, 2019}
\accepted{January 31, 2019}
\submitjournal{ApJ}
\shorttitle{Radio jet properties of 3C\,411}
\shortauthors{Perger et al.}

\begin{document}

\title{Is there a Blazar Nested in the Core of the Radio Galaxy 3C\,411?}

\correspondingauthor{Krisztina Perger}
\email{k.perger@astro.elte.hu}

\author[0000-0002-6044-6069]{Krisztina Perger}
\affiliation{Department of Astronomy, E\"otv\"os Lor\'and University \\
P\'azm\'any P\'eter s\'et\'any 1/A,
H-1117 Budapest, Hungary}
\affiliation{Konkoly Observatory, MTA Research Centre for Astronomy and Earth Sciences\\
 Konkoly Thege Mikl\'os \'ut 15-17, H-1121 Budapest, Hungary}

\author[0000-0003-3079-1889]{S\'andor Frey}
\affiliation{Konkoly Observatory, MTA Research Centre for Astronomy and Earth Sciences\\
 Konkoly Thege Mikl\'os \'ut 15-17, H-1121 Budapest, Hungary}

\author[0000-0003-1020-1597]{Krisztina \'E. Gab\'anyi}
\affiliation{MTA-ELTE Extragalactic Astrophysics Research Group, E\"otv\"os Lor\'and University\\ P\'azm\'any P\'eter s\'et\'any 1/A,
 H-1117 Budapest, Hungary}
\affiliation{Konkoly Observatory, MTA Research Centre for Astronomy and Earth Sciences\\
 Konkoly Thege Mikl\'os \'ut 15-17, H-1121 Budapest, Hungary}

%--------------------------------------------------------------------------------
%--------------------------------------------------------------------------------
%-------------------------------ABSTRACT-----------------------------------------
%--------------------------------------------------------------------------------
%--------------------------------------------------------------------------------

\begin{abstract}
Previous spectral energy distribution modeling based on \textit{XMM-Newton} X-ray observation of the classical double-lobed radio galaxy 3C\,411 left the possibility open for the presence of a blazar-like core. We investigated this scenario by characterizing the radio brightness distribution in the inner $\sim$10-pc region of the source. We applied the very long baseline interferometry (VLBI) technique at four different frequencies from 1.7 to 7.6~GHz. We analyzed archival data from the Very Long Baseline Array (VLBA) taken in 2014, and data from the European VLBI Network (EVN) obtained in 2017. The VLBI images reveal pc-scale extended structure in the core of 3C\,411 that can be modeled with multiple jet components. The measured core brightness temperatures indicate no Doppler enhancement that would be expected from a blazar jet pointing close to the line of sight. While there is no blazar-type core in 3C\,411, we found indication of flux density variability. The overall morphology of the source is consistent with a straight jet with $\sim50\degr$ inclination angle at all scales from pc to kpc.
\end{abstract}


%--------------------------------------------------------------------------------
%--------------------------------------------------------------------------------
%----------------------------------INTRO-----------------------------------------
%--------------------------------------------------------------------------------
%--------------------------------------------------------------------------------

\section{Introduction}\label{sec:intro}
The primary classification of active galactic nuclei (AGN) is based on the spatial orientation of their axis of symmetry with respect to the line of sight to the observer. It is called the unified model of AGN \citep[e.g.][]{antonucci1993,urry1995,netzer2015}. The orientation is imprinted in the optical spectral lines of the galactic cores. Those with small viewing angle show both broad permitted and narrow forbidden emission lines in their spectra (type 1). Those oriented at a large angle to the line of sight have only narrow emission lines due to the obscured broad line forming region by a dusty feature around the central engine (i.e. an accreting supermassive black hole, SMBH) and the origin of the broad line forming region (type 2). Nearly 10\% of AGN have relativistic plasma jets ejected in two opposite directions from the vicinity of the central SMBH along its rotation axis. These jets produce radio emission via synchrotron mechanism. 
In the case of type 1 jetted AGN, the small viewing angle might cause the outflowing material appear moving at superluminal speeds \citep[e.g.][]{rees1966} and the intensity of the radiation originating from the approaching jet is significantly enhanced by Doppler boosting \citep{shklovskii1964}.

The earliest studies \citep[e.g.][]{spinrad1975} classified the target 3C\,411 (right ascension $20^\mathrm{h} 22^\mathrm{m} 08\fs440$, declination $+10\degr 01\arcmin 11\farcs314$) as an AGN by its bright optical nucleus and strong emission in the radio regime detected in the third Cambridge (3C) survey \citep{edge1959}. Broad emission lines in its optical spectrum led to the classification of the source as a type 1 Seyfert galaxy  \citep{khachikian1974,veron-cetty2006}. The radio counterpart was first resolved into a double-lobed structure extending to $\sim 26\arcsec$ with interferometric measurements by \citet{spinrad1975}. The large-scale radio structure of 3C\,411 consists of a pair of approximately symmetrical hot spots and lobes dominating the radio emission, and a significantly fainter core centered on the optical AGN. Based on its radio morphology, 3C\,411 is a Fanaroff--Riley class II (FR\,II) radio galaxy \citep{fanaroff1974}. Very Large Array (VLA) measurements at 1.4~GHz by \citet{spangler1984} gave an intensity ratio of the two hot spots $K=2.35$  in favor of the south-eastern one, and indicated the presence of a bridge-like emission at the western side of the core with an absent jet. Flux densities $28\pm2$~mJy and $45\pm2$~mJy were also determined for the core component at 1.4 and 5~GHz, respectively, resulting in a spectral index of the inner structure of $ \alpha_{1.4,\mathrm{in}}^{5}=0.39$ \citep{spangler1984} by the convention of $S_\nu\propto\nu^{\alpha}$ (where $\nu$ is the frequency and $S_\nu$ is the flux density).

The classification of 3C\,411 within the framework of the unified model of AGN was challenged by \citet{bostrom2014} who observed the source in X-rays with the \textit{XMM-Newton} satellite. They analyzed the spectral energy distribution of the source in the energy range of $0.2-10$~keV by fitting several spectral models: a simple power-law, an absorbed power-law with a blackbody component (accretion disc), an absorbed double power-law (accretion disc with reflection spectrum), a complex accretion disc model called `kdblur' (Galactic absorption, disc inclination and iron abundance), and two models that took the presence of warm absorbers into account. Two models among those listed above provided equally satisfactory fit to the data: the `kdblur' and the double power-law models. The disc-dominated `kdblur' model is consistent with the current classification of 3C\,411 as an AGN lying close to the plane of the sky (i.e. misaligned). On the contrary, the double power-law model implies a jet-dominated system emitting a Seyfert-like softer and a blazar-like harder X-ray component, essentially requiring that the nucleus of 3C\,411 is a blazar. This would imply a major difference in the direction of the relativistic jet between the innermost pc-scale and the outer kpc-scale regions of the source, deviating from the general picture of the unified model. If this is indeed the case, together with a notable sample of AGN reported with similar dichotomy \citep[e.g.][and references therein]{kharb2010}, 3C\,411 could provide new details to refine the unified model. 

Here we invoke high-resolution radio observations using the technique of very long baseline interferometry (VLBI) to characterize the compact radio emission of the core of 3C\,411 (PKS\,J2022+1001). The VLBI observations are capable of providing clear direct evidence for the blazar nature of the core by means of detecting Doppler-boosted radio emission, therefore we can distinguish between the two competing models of the X-ray emission \citep{bostrom2014}. We use new Europen VLBI Network (EVN) observations at 1.7 and 5~GHz and archival Very Long Baseline Array (VLBA) snapshots at 4.3 and 7.6~GHz.

In Section~\ref{sec:data}, we introduce the observing conditions and the methods of data reduction. 
Section~\ref{sec:results} presents the results of our data analysis. These are discussed in Section~\ref{sec:discussion}. Concluding remarks and a summary are given in Section~\ref{sec:conclusions}.

We adopt $\Lambda$CDM cosmological model parameters $H_0=70$~km~s$^{-1}$~Mpc$^{-1}$, $\Omega_\mathrm{M}=0.3$, and $\Omega_\Lambda=0.7$. At the redshift of 3C\,411 \citep[$z=0.467$;][]{spinrad1975}, 1 milliarcsecond (mas) angular size corresponds to 5.88~pc projected linear size \citep{wright2006} in this model.

%--------------------------------------------------------------------------------
%--------------------------------------------------------------------------------
%----------------------------------------DATA------------------------------------
%--------------------------------------------------------------------------------
%--------------------------------------------------------------------------------

\begin{table*}[ht!]
\centering
\caption{Details of the observations}
\begin{tabular}{ccccl}
 Observation date 		&On-source time					& Array			& $\nu$ (GHz)	& Participating telescopes\\\hline\hline
 2014 February 3		&$39^\mathrm{s}$				& VLBA			& 4.3				& BR, FD, HN, KP, LA, NL, OV, PT, SC \\
 2014 February 3		&$39^\mathrm{s}$				& VLBA			& 7.6				& BR, FD, HN, KP, LA, NL, OV, PT, SC \\
 2017 May 31			&$1^\mathrm{h}40^\mathrm{m}$	& EVN			& 1.7				&  BD, EF, HH, IR, JB, MC, SV, T6, TR, UR, WB, ZC\\
 2017 June 12			&$1^\mathrm{h}40^\mathrm{m}$	& EVN			& 5.0				&  BD, EF, HH, IR, JB, MC, NT, SV, T6, TR, UR, WB, YS, ZC\\
\hline
\end{tabular}

\tablecomments{BR -- Brewster (Washington), FD -- Fort Davis (Texas), HN -- Hancock (New Hampshire), KP -- Kitt Peak (Arizona), LA -- Los Alamos (New Mexico), NL -- North Liberty (Iowa), OV -- Owens Valley (California), PT -- Pie Town (New Mexico), SC -- St. Croix (U.S. Virgin Islands); BD -- Badary (Russia), EF -- Effelsberg (Germany), HH --Hartebeesthoek (South Africa), IR -- Irbene (Latvia), JB -- Jodrell Bank Mk2 (United Kingdom), MC -- Medicina (Italy), NT -- Noto (Italy), YS -- Yebes (Spain), SV -- Svetloe (Russia), T6 -- Tianma (China), TR -- Toru\'n (Poland), UR -- Nanshan (China), WB -- Westerbork (the Netherlands), ZC -- Zelenchukskaya (Russia)}\label{tab:obs}
\end{table*}

\section{Observations and data reduction} 
\label{sec:data}

The EVN observations (project code: EP104, PI: K. Perger) were done on 2017 May 31 and June 1 at 1.7~GHz, and on 2017 June 12 at 5~GHz. Both project segments lasted for a total of 2~h. The data were recorded in left and right circular polarizations at 1024 Mbit~s$^{-1}$ data rate. Eight intermediate frequency channels (IFs) were used, with the total bandwidth of 128~MHz per polarization in 32 spectral channels per IF. Details of the observations including the participating radio telescopes are listed in Table~\ref{tab:obs}. The data were correlated at the Joint Institute for VLBI ERIC (Dwingeloo, the Netherlands) with 2~s averaging time for both the 1.7 and 5~GHz observations.

We calibrated the data in the US National Radio Astronomy Observatory (NRAO) Astronomical Image Processing System\footnote{http://www.aips.nrao.edu/index.shtml} \citep[e.g.][]{diamond1995}. After a priori amplitude calibration that was based on the antenna system temperatures and gain curves, we performed manual phase calibration on a 1-min data segment of a bright calibrator source (J2123+0535). After inspecting the data and flagging, global fringe-fitting was performed on the target source 3C\,411 data. Then we exported them into the {\sc difmap} program for imaging \citep{shepherd1997} and model-fitting. The baselines to Irbene were excluded from the analysis due to amplitude calibration problems at 1.7~GHz.

The VLBA observations of 3C\,411 were carried out on 2014 February 3 in the framework of the 8th VLBA Calibrator Survey (project code: BP177F, PI: L. Petrov). Nine antennas of the array were involved (see Table~\ref{tab:obs}). Eight IFs were used with a total bandwidth of 256~MHz in right circular polarization. The calibrated \textsc{uvfits} data files created by the \textsc{PIMA} v2.08 software \citep{petrov2011} were obtained from the Astrogeo VLBI database\footnote{http://astrogeo.org/}.

\begin{figure*}[ht!]
\centering
\includegraphics[width=0.8\linewidth]{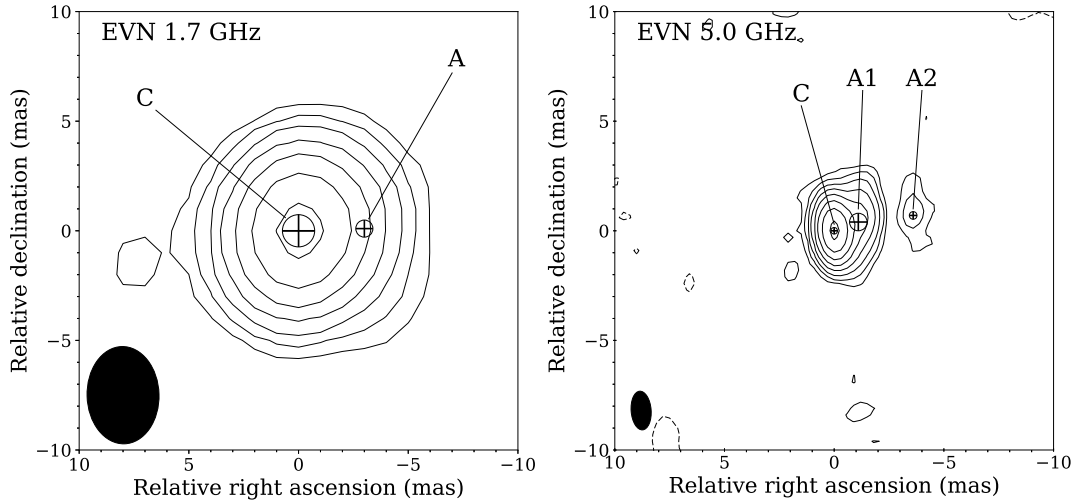}
\caption{Naturally weighted clean maps of 3C\,411 at 1.7~GHz (left) and 5~GHz (right) from the EVN observations. Contour levels start at $\pm0.21$ and $\pm0.18$~mJy~beam$^{-1}$, positive levels increase by a factor of 2, peak intensities are $16.9$ and $27.3$~mJy~beam$^{-1}$ for the 1.7 and 5~GHz image, respectively. The Gaussian restoring beam is shown in the bottom left corners. Sizes and positions of the fitted circular Gaussian model components are indicated with crossed circles, components are labeled corresponding to Table~\ref{tab:model}.}\label{fig:evn}
\end{figure*}

\begin{figure*}[ht!]
\centering
\includegraphics[width=0.8\linewidth]{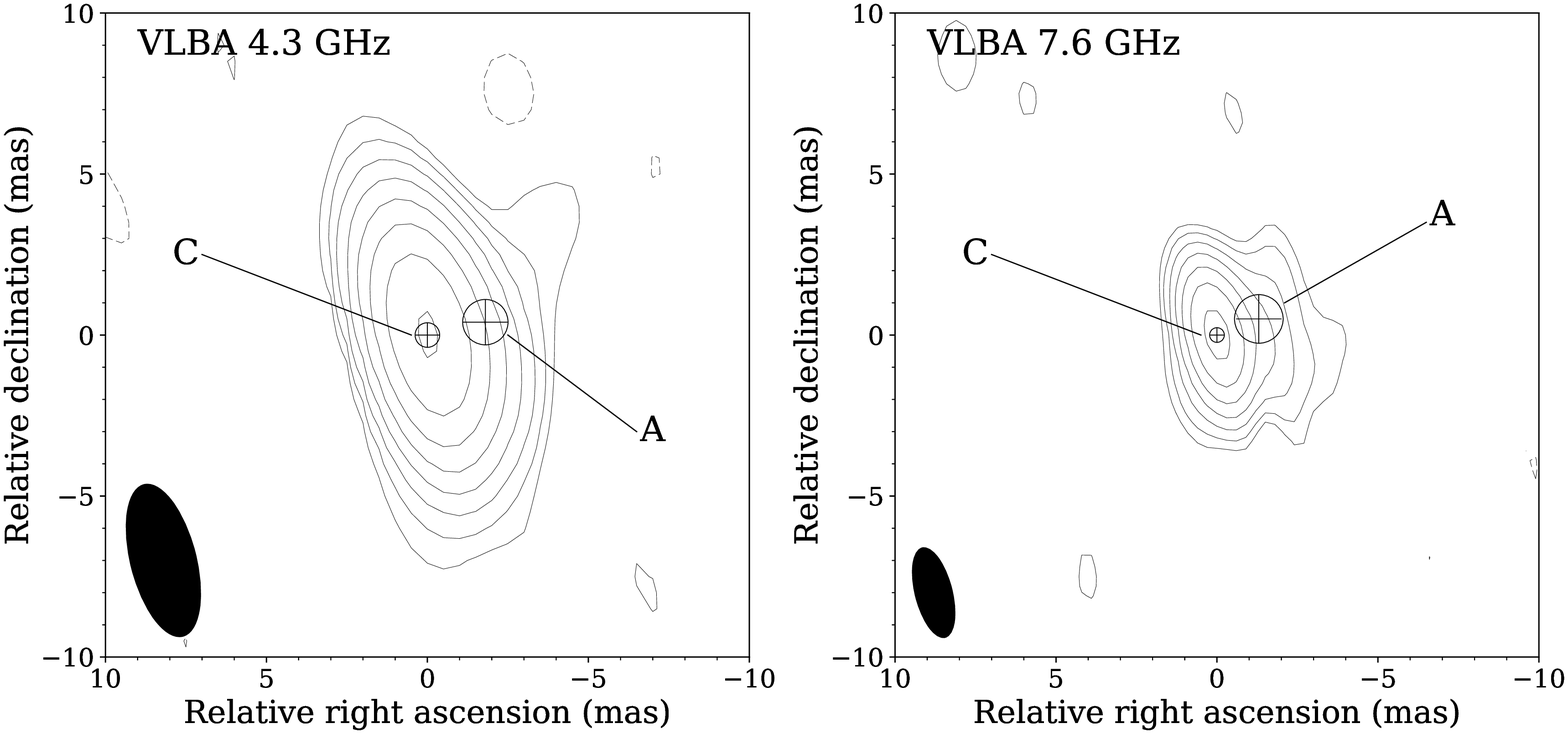}
\caption{Naturally weighted clean maps of 3C\,411 at 4.3~GHz (left) and 7.6~GHz (right) from the VLBA observations. Contour levels start at $\pm0.3$ and $\pm0.4$~mJy~beam$^{-1}$, positive levels increase by a factor of 2, peak intensities are $41.4$ and $32.6$~mJy~beam$^{-1}$ for the 4.3 and 7.6~GHz image, respectively. The Gaussian restoring beam is shown in the bottom left corners. Sizes and positions of the fitted circular Gaussian model components are indicated with crossed circles, components are labeled corresponding to Table~\ref{tab:model}.}\label{fig:vlba}

\end{figure*}

For both the EVN and VLBA data, we used the conventional hybrid mapping procedure in {\sc difmap} involving cycles of {\sc clean} deconvolution \citep{hogbom1974} and phase self-calibration. At the final stage, amplitude self-calibration was also applied. The EVN and VLBA contour images are displayed in Figs.~\ref{fig:evn} and \ref{fig:vlba}, respectively. The self-calibrated visibility data were then fitted with multiple circular Gaussian model components \citep{pearson1995}, to quantitatively characterize the source brightness distribution. Model fitting parameters are listed in Table~\ref{tab:model}. The estimated errors for model parameters were calculated following \citet{lee2008}, considering an additional $5\%$ amplitude calibration uncertainty in the flux densities.

%--------------------------------------------------------------------------------
%--------------------------------------------------------------------------------
%-------------------------------------RESULTS------------------------------------
%--------------------------------------------------------------------------------
%--------------------------------------------------------------------------------

\begin{table*}[ht!]
\centering
\caption{Image and model parameters}\label{tab:model}
\begin{tabular}{ccrrrrccc}
$\nu$ (GHz)& Comp.  & $S$ (mJy)				& $\vartheta$ (mas) & $R$ (mas)			&$\phi$ ($^\circ$)		&$T_\mathrm{B}$ ($10^{10}$~K)&$\delta_\mathrm{eq}$&$\delta_*$\\

\hline
\hline
1.7 	& C			&$16.0 \pm 1.6$		&$0.82\pm 0.05$		&0					&0						&$1.5 \pm 0.3$	    &$0.31 $ &$ 0.55$\\
	 	& A			&$4.1 \pm 0.7$		&$<0.45  \pm 0.06$		&$2.08 \pm 0.03$	& $-78.2 \pm 0.9 $ 		&	--			&--&--\\

4.3		& C			&$38.3 \pm 4.5 $		&$<0.36\pm  0.03$	&0					&0						&$2.7\pm 0.7$	&$0.55 $&$ 0.91$\\
	 	& A			&$8.6  \pm 1.4 $		&$<0.81  \pm 0.13 $	&$1.37 \pm 0.06$	& $-78.9\pm3.1$ 		&	--			&--&--\\

5.0 	& C			&$28.7 \pm 2.2$		 &$0.28 \pm 0.01 $		&0					&0						&$2.6 \pm 0.4$ &0$.53 $&$ 0.88$\\
		& A1			&$6.4 \pm 0.9$ 		&$0.74 \pm 0.08 $		&$1.11 \pm 0.04$	&$-68.6 \pm 2.2 $	&	--			&--&--\\
		& A2			&$0.9 \pm 0.3$ 		&$<0.30  \pm 0.07$		&$3.52 \pm 0.04$ 	&$-78.4 \pm 0.6 $	&	--			&--&--\\

7.6 	& C			&$ 36.3\pm 4.1 $ 		&$0.48  \pm  0.04$	&0					&0						&$0.5 \pm 0.1$	&$0.11$ &$ 0.18$\\
		& A			&$4.0  \pm1.2  $		&$<0.40 \pm  0.09$	&$1.22\pm 0.04$		&$-69.6\pm10.9 $		&	--			& --&--\\

\hline
\end{tabular}
\tablecomments{ Column 1 -- observing frequency, Column 2 -- model component name, Column 3 -- flux density, Column 4 -- circular Gaussian model component size (FWHM) or upper limit corresponding to the minimum resolvable angular size \citep{kovalev05}, Column 5 -- radial angular distance from the core, Column 6 -- position angle with respect to the core, measured from north through east, Column 7 --  brightness temperature of the core component,  Column 8 -- Equipartition Doppler factor \citep{readhead1994}, Column 9 -- Doppler factor assuming $T_\mathrm{b,int}=3\times10^{10}$~K \citep{homan2006}}
\end{table*}

\section{Results}\label{sec:results}

\subsection{Inner jet structure}

The image of 3C\,411 made at the lowest frequency, 1.7~GHz (left panel in Fig.~\ref{fig:evn}) shows a barely resolved compact core. However, the best-fit model (Table~\ref{tab:model}) consists of two circular Gaussian components that reveal the substructure in the jet pointing towards the west. The location and the full width at half-maximum (FWHM) size of the components are also indicated in the image. The jet component is labeled as A.

Moderate structural details are visible also in the 4.3~GHz VLBA map, but the jet starts to be better resolved at 7.6~GHz (Fig.~\ref{fig:vlba}). According to model fitting, apart from the core, a second component (labeled as A) is detected at both frequencies.  These observations were made simultaneously.

The finest angular resolution was provided by the EVN observation at 5~GHz. The image (right panel in Fig.~\ref{fig:evn}) shows a well resolved structure with two `jet' components (A1 and A2) within 4~mas from the core (C).

We note that the 1.7 GHz interferometric visibility phases and amplitudes as a function of time displayed a distinct modulation on the shortest Effelsberg--Westerbork baseline. The fringe spacing corresponding to the $B=266$~km long baseline is $\lambda / B \approx 140$~mas where $\lambda$ is the wavelength. We interpret this modulation as a contribution of the two hot spots in the large-scale radio structure of 3C\,411. However, these are at $\sim13\arcsec$ angular distance, beyond the undistorted field of view in the EVN observations. With the addition of extra model components at approximate positions of the hot spots known from VLA imaging \citep{spangler1984,neff1995}, we could reproduce the time variability of the phases on this short baseline. This indicates that the hot spots are compact at $\sim0\farcs1$ level.

\subsection{Jet parameters}

We calculated the brightness temperatures for the bright central core components (C) at all four frequency bands \citep[e.g.][]{condon1982}:
\begin{equation}\label{eq:tb}
T_\mathrm{b}=1.22 \times 10^{12} \, (1+z) \, \frac{S}{\vartheta^2\nu^2}\,\text{K}.
\end{equation}
Here $S$ is the flux density in Jy, $\vartheta$ is the size of the circular Gaussian component (FWHM) in mas, and $\nu$ is the observing frequency in GHz. The results are listed in Table~\ref{tab:model}. 

With a reasonable assumption about the intrinsic brightness temperature $T_\mathrm{b,int}$, we can determine if the jet radio emission is enhanced by Doppler boosting. The Doppler factor is
\begin{equation}\label{eq:doppler}
\delta=\frac{T_\mathrm{b}}{T_\mathrm{b,int}}.
\end{equation}
For the intrinsic brightness temperature, we considered {\bf a} larger and a smaller estimate used in the literature. The former value, $T_\mathrm{b,int} \approx 5\times 10^{10}$~K corresponds to the energy equipartition between the radiating particles and the magnetic field \citep{readhead1994}. A somewhat lower value, $T_\mathrm{b,int} \approx 3 \times 10^{10}$~K was derived by \citet{homan2006} based on a study of a sample of pc-scale AGN jets. In Table~\ref{tab:model}, we list both Doppler factors in col. 8 and col. 9. Note that a high correlation between the inverse Compton and equipartition Doppler factors were found by \citet{guijosa1996}, and Doppler factors of radio galaxies like 3C\,411 are lower than for other radio AGN.

\subsection{Radio spectrum}\label{sec:spectrum}

\begin{figure}[htbp]
\centering
\includegraphics[width=\linewidth]{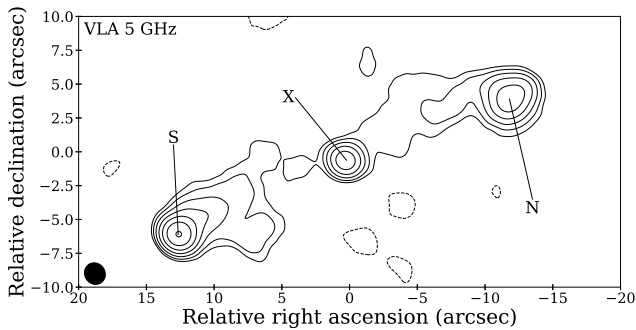}
\caption{The VLA image of 3C\,411 at 5~GHz \citep{neff1995}.  The first contours are drawn at $\pm1.5\%$ of the peak intensity ($220.4$~mJy~beam$^{-1}$), the levels scale up by a factor of 2. The size of the restoring beam is $1\farcs5\times1\farcs63$. The central radio source, the subject of our VLBI study, is marked with X. The south-eastern and north-western complexes are denoted by S and N, respectively.\label{fig:vla}}
\end{figure}

\begin{figure}[ht]
\centering
\includegraphics[width=\linewidth]{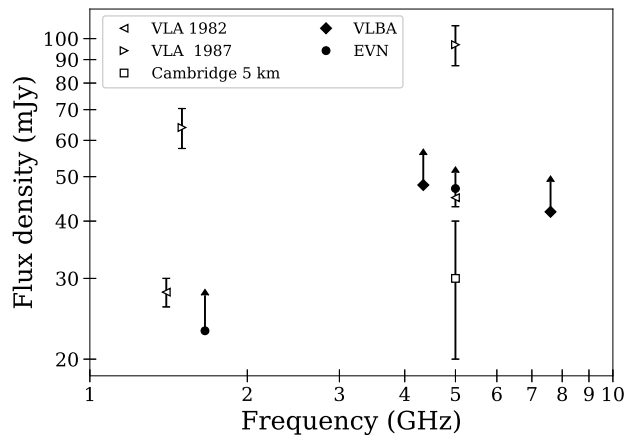}%spektrum_korlat.eps
\caption{Non-simultaneous radio spectrum of the central component in 3C~411. Because of the high resolution compared to that of the connected-element interferometers and thus the possibility that some arcsec-scale emission is resolved out, data points from the VLBI observations presented here are shown as lower limits with arrows.}  

\label{fig:spectrum}
\end{figure}

\begin{figure}
\centering
    \includegraphics[width=\linewidth]{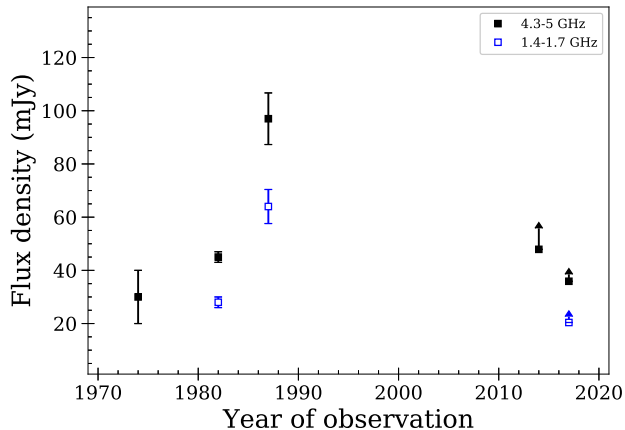}
    \caption{Radio lightcurve of the centre of 3C~411. Flux densities at $\sim$1.4 and $\sim$5~GHz reveal the variable nature of the AGN. VLBI measurements are shown as lower limits with errors. }
    \label{fig:lightcurve}
\end{figure}

We compiled the radio spectrum of the central source in 3C~411 (labeled as X in Fig.~\ref{fig:vla}) from the literature \citep{pooley1974,spangler1984,neff1995} and from our own measurements (Fig.~\ref{fig:spectrum}). VLBI flux densities are considered as lower limits because part of the radio structure that is extended to arcsec scale may be resolved out. Multi-epoch flux density measurements at around 1.4 and 5~GHz clearly show variability (Fig.~\ref{fig:lightcurve}). 

 We calculated the two-point spectral index of the source separately from the VLBA and EVN observations and  found   $\alpha_{5,\mathrm{EVN}}^{1.7}=0.51\pm0.05$ and $\alpha_{7.6,\mathrm{VLBA}}^{4.3}=-0.26\pm0.06$. The 5-to-1.7~GHz EVN spectral index is slightly steeper than those calculated from VLA data at the same frequencies (for component X in Fig.~\ref{fig:vla}), $0.39\pm0.04$ and $0.35\pm0.06$ for the 1982 and 1987 observations, respectively \citep{spangler1984, neff1995}. This can be attributed to the variability of the source, or the different resolution. The spectral indices for EVN and VLA data, and the overall shape of the spectrum suggest that it is peaking at $\sim 5$~GHz. Simultaneous multi-frequency flux density measurements would be needed for a reliable determination of the turnover frequency.

%--------------------------------------------------------------------------------
%--------------------------------------------------------------------------------
%---------------------------------DISCUSSION-------------------------------------
%--------------------------------------------------------------------------------
%--------------------------------------------------------------------------------

\section{Discussion}\label{sec:discussion}

\subsection{Inner structure: is it a blazar jet?}

The compact mas-scale jet structure of the central radio source in 3C\,411 (Figs.~\ref{fig:evn} and \ref{fig:vlba}) is typical of radio quasars, though  one-sided jets are also present at pc scales in a number of radio galaxy cores \citep[][and references therein]{giovannini2001}. Apart from the synchrotron self-absorbed base of the jet (i.e. the core C), the jet extending to the western direction is decomposed into multiple components, depending on the observing frequency and the angular resolution of the interferometer. Component A identified with model fitting to the VLBA data at 4.3 and 7.6~GHz, and to the EVN data at 1.7~GHz (Figs.~\ref{fig:evn} and \ref{fig:vlba}, Table~\ref{tab:model}) is apparently further resolved into two distinct features by the highest-resolution 5-GHz EVN observations (Fig.~\ref{fig:evn}, right). In principle, because of the 3.32 yr time difference between the VLBA and EVN observations, component A could have been propagated into the position of A2. In this scenario, component A1 on the EVN maps would be a newly-emerged blob in the jet appearing some time after 2014 February, the epoch of the VLBA observations. However, both component movements would require unphysically high bulk Lorentz factors ($\Gamma > 250$) that have never been seen in AGN. In fact, VLBI studies of pc-scale radio jets \citep[e.g.][]{kellermann2004,lister2016,pushkarev2017} found that Lorentz factors have an upper limit of $\Gamma \approx 30$. Positional differences of the 4.3 and 7.6~GHz VLBA `jets' can be explained as component A (Table~\ref{tab:model}) in the 4.3~GHz data is the combination of A1 and A2 at the 5 GHz EVN maps, however at 7.6 GHz, the `outer'  component is too faint or diffuse to contribute to the overall emission of A.

The Doppler factors derived from our VLBI measurements (Table~\ref{tab:model}) are all below unity. Therefore the radio emission is Doppler-deboosted, contrary to the expectations from a blazar jet. There is no widely accepted definition of how small the viewing angle ($\theta$) of blazar jets should be, but we can adopt $\theta < 1/\Gamma$ \citep{ghisellini2016}, where $\Gamma$ is the bulk Lorentz factor of the outflowing plasma. Below we estimate the jet viewing angle in the central region of 3C\,411.   

According to Table~\ref{tab:model}, the determined Doppler factors at all frequencies but 7.6~GHz are rather similar, $0.31\la \delta \la 0.98$. Since equipartition Doppler factors calculated at the peak of the spectrum approach the real Doppler factor the  best \citep{readhead1994},  we consider the other Doppler factor estimates as lower limits and use the range $0.53 \le \delta \le 0.91$ determined at 4.3 and 5~GHz in the subsequent calculations. Also, the resolution of the EVN array is the best at 5~GHz where the innermost mas-scale jet features are not blended, and for which the Doppler factor range coincides with that of the 4.3~GHz VLBA data.

\begin{figure}[h]
\centering
\includegraphics[width=\linewidth]{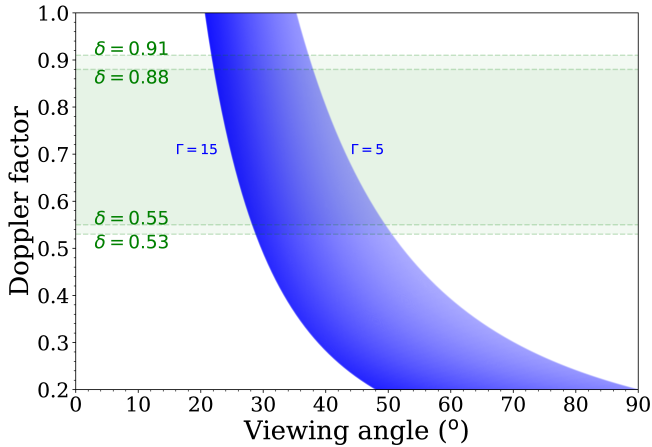}
\caption{Doppler factors versus viewing angles in the Doppler-deboosted regime ($\delta < 1$). The green areas denote the calculated Doppler factor ranges at around the spectral turnover, 4.3 and 5~GHz (Table~\ref{tab:model}), the blue area indicates the assumed range of Lorentz factors.}\label{fig:doppler}
\end{figure}

Because multi-epoch VLBI monitoring data at a fixed frequency that would be sufficient for determining the apparent speed of any jet component in 3C\,411 are not available, the Lorentz factor cannot be constrained. Instead, we assume reasonable $\Gamma$ values obtained from the literature. Parsec-scale jets in radio galaxies similar to our source populate the $5 \la \Gamma \la 15$ regime \citep{kellermann2004,cohen2007,baldi2013}. Using the formulae for relativistic beaming parameters \citep[see e.g. Appendix A in][]{urry1995}, we plot the Doppler factor as a function of jet viewing angle, marking the assumed range of Lorentz factors in Fig.~\ref{fig:doppler}. For $\Gamma=15$, the possible viewing angles exceed $\sim 22\degr$. Lower Lorentz factors imply even larger $\theta$ values. Obviously the $\theta < 1/\Gamma$ criterion is not fulfilled and therefore 3C\,411 does not host a blazar in its center. However, we can conclude that the VLBI data indicate an AGN nucleus with jet  inclination angle $ 20\degr \la \theta \la 50\degr$ (Fig.~\ref{fig:doppler}). 

\subsection{Large-scale radio structure}

To better understand the relation between the pc- and kpc-scale radio structure of 3C\,411, we utilize the results of \citet{spangler1984}, and reproduce (Fig.~\ref{fig:vla}) and analyze the 5-GHz VLA image made by \citet{neff1995} which we obtained from the NASA/IPAC Extragalactic Database (NED)\footnote{https://ned.ipac.caltech.edu}.

Based on these images, we identify the approaching and receding sides of the large-scale jets and estimate their inclination angle with respect to the line of sight. The total extent of the source corresponds to a projected linear size of $155$~kpc.

In the following, we assume that the jet activity started simultaneously on both sides of the central engine and the expansion took place in an intrinsically similar way in the two opposite directions. The apparent asymmetry in intensity is therefore caused by relativistic beaming and the radiation coming from the more distant counterjet side travels longer through the interstellar medium until it reaches the observer.  According to the Laing--Garrington effect \citep{garrington1988,laing1988}, the lobe associated with the approaching jet is usually brighter and shows less depolarization. Also, the brighter to the fainter lobe arm-length ratio is larger than one \citep{longair1979}. 

In the case of 3C\,411 the  southeastern (S) hot spot is brighter, more compact \citep{spangler1984, neff1995}, and has a slightly longer projected distance from the central component (X) than  the northwestern (N) one ($\phi_\mathrm{NX}=12\farcs9 \pm 0\farcs3$ and $\phi_\mathrm{SX}=13\farcs5 \pm 0\farcs3$, Fig.~\ref{fig:vla}). This corresponds to an arm-length ratio $R=1.05\pm0.03$, what would imply that the radio galaxy lies close to the plane of thy sky.  \citet{spangler1984} made a detailed study on the polarization properties of the AGN, and they found a higher depolarization parameter in S ($15\pm1$) than in N ($\sim3.5$), suggesting N being on the approaching side. The brightness, compactness, and projected distance of the hot spots from the central source are in favour of S being the approaching side,  however, according to \citet{hardcastle1998}, who studied a large sample of FR~II radio galaxies, there is no relationship between the jet sidedness and the brightness of the hot spots. Considering all above, and especially the presence of a kpc-scale jet visible on the northwestern side of the source, \citep[see fig.~3 in][]{spangler1984}, we conclude that N is on the approaching and S is on the receding side. 

In the following we estimate the inclination angle using the kpc-scale structure, the $\sim2$~arcsec jet in component X, with the help of the formula \citep{saikia1989, taylor1997}
\begin{equation}
K=\left(\frac{1+\beta\cos\theta}{1-\beta\cos\theta}\right)^{2-\alpha},
\end{equation}
where $\beta=vc^{-1}$ is the speed of the jet, $K$ is the flux density ratio between the approaching and receding sides,  and $\alpha$ is the spectral index. The absence of the counterjet on arcsec scales (in the central source, X) leads to a ratio $K\geq9$. With $\alpha=\alpha_{1.4,\mathrm{in}}^{5}=0.39$ \citep{spangler1984} and $\beta<1$,  an upper limit can be derived on the kpc-scale inner jet inclination angle $\theta\la50\degr$.

The apparent discrepancy between the inclination angle determined in the inner kpc-scale and the estimation on the outer kpc-scale arm-length could be that the intrinsic symmetry between the opposite jet sides is not a valid assumption at all regions. It is possible that asymmetric conditions in the ambient interstellar medium cause different deceleration of the jet on the two sides. 

The estimated angles to the line of sight are comparable both at the inner kpc and pc scales of the radio structure. Similarly, as projected on the plane of the sky,  the VLBI  jet model component position angle at 5~GHz ($-69\degr\pm2\degr$) coincides with the position angle inferred from the relative positions of the N and S hot spots (outer kpc-scale), about $-67\degr$, indicating  no significant bending in the overall structure of the jet.

%--------------------------------------------------------------------------------
%--------------------------------------------------------------------------------
%----------------------------------CONCLUSIONS-----------------------------------
%--------------------------------------------------------------------------------
%--------------------------------------------------------------------------------

\section{Summary and conclusions}\label{sec:conclusions}
We used new EVN observations and archival VLBA measurements to describe the pc-scale radio structure of the radio galaxy 3C\,411, supplemented by archival VLA data for kpc scales. With the aim of confirming or rejecting the blazar presence in the AGN core, we fitted circular Gaussian model components to the VLBI data and calculated the core brightness temperatures. The inferred Doppler boosting factors are smaller than unity, providing no evidence for a blazar core hidden in the center of 3C\,411. Therefore we ruled out one of the alternative models put forward by \citet{bostrom2014} to explain the X-ray spectral energy distribution of 3C\,411.

 Based on an archival 5-GHz VLA image of the source \citep{neff1995}, we estimated the inclination angle of the kpc-scale jet which is in agreement with the value obtained from the VLBI data. Therefore the radio structure of 3C\,411 is consistent with an oppositely-directed straight jet launched from the central engine, all the way from pc to $\sim 100$~kpc scales.

%--------------------------------------------------------------------------------
%--------------------------------------------------------------------------------
%----------------------------------ACKNOWLEDGMENTS-------------------------------
%--------------------------------------------------------------------------------
%--------------------------------------------------------------------------------

\acknowledgments
We thank the anonymous referee for his/her valuable suggestions that improved the discussion.
The EVN is a joint facility of independent European, African, Asian and North American radio astronomy institutes. Scientific results from data presented in this publication are derived from the following EVN project code: EP104. The NRAO is a facility of the National Science Foundation operated under cooperative agreement by Associated Universities, Inc. This publication made use of the Astrogeo VLBI FITS image database (http://astrogeo.org/vlbi$\_$images/) and the authors thank Leonid Petrov for making results of 3C411 observations online prior to publication. We thank the Hungarian National Research, Development and Innovation Office (OTKA NN110333) for support. K\'{E}G was supported by the J\'{a}nos Bolyai Research Scholarship of the Hungarian Academy of Sciences. This publication has received funding from the European Union's Horizon 2020 research and innovation programme under grant agreement No 730562 (RadioNet). This research has made use of the NASA/IPAC Extragalactic Database (NED) which is operated by the Jet Propulsion Laboratory, California Institute of Technology, under contract with the National Aeronautics and Space Administration.

\software{AIPS \citep{greisen2003}, Difmap \citep{shepherd1997}, Astropy \citep{astropy2013}, Matplotlib \citep{hunter2007}}

%--------------------------------------------------------------------------------
%--------------------------------------------------------------------------------
%----------------------------------BIBLIOGRAPHY----------------------------------
%--------------------------------------------------------------------------------
%--------------------------------------------------------------------------------

\end{document}